\documentstyle[twoside,fleqn,epsf,espcrc2]{article}

\newcommand{\plus}{\makebox[15pt][c]{$+$}}
\newcommand{\minus}{\makebox[15pt][c]{$-$}}

\newcommand{\err}[2]
{\hskip-0.5em\raisebox{0.08em}{\scriptsize{$\;\begin{array}{@{}l@{}}
\plus\makebox[0.50em][r]{#1\hfill} \\[-0.24em]
\minus\makebox[0.50em][r]{#2\hfill} 
\end{array}$}}}

\newcommand{\AmS}{{\protect\the\textfont2
  A\kern-.1667em\lower.5ex\hbox{M}\kern-.125emS}}
 
\title{$\bar{B}\longrightarrow\pi l\bar{\nu}$ at three lattice spacings
 \hfill\normalsize FERMILAB-CONF-98/269-T
}

\author{
S.~Ryan$^a$\thanks{talk presented by S.~Ryan.}, A.~El-Khadra$^b$, S.~Hashimoto$^a$, A.~Kronfeld$^a$, P.~Mackenzie$^a$ and J.~Simone$^a$\\[5mm]
$^a$ {\small Theoretical Physics Department, Fermilab, P.O. Box 500, Batavia, Il 60510, U.S.A.}\\
$^b$ {\small Loomis Laboratory of Physics, 1110 W. Green Street, Urbana, Il 61801-3080, U.S.A.} 
}
\begin{document}  
\begin{abstract}
The increasing accuracy of experimental results for the exclusive, semileptonic decay $\bar{B}\longrightarrow \pi l\bar{\nu}$ requires a similarly accurate calculation of the hadronic matrix elements, to determine $|V_{ub}|$.
We present preliminary results for the form factors of the $B$ to light meson decay mode.
Using results from three lattices in the range $5.7 \leq \beta \leq 6.1$ we study the dependence on the lattice spacing.

\end{abstract}
\maketitle
\section{Introduction}
In this report we present preliminary results from our study of semileptonic $\bar{B}\rightarrow\pi l\bar{\nu}$ decays. This exclusive decay mode can be used to extract the CKM matrix element $|V_{ub}|$ which is currently known to only $\sim 20\%$ accuracy. The experimental error will be greatly reduced at $B$-factories, requiring a similar reduction in the theoretical error to place meaningful constraints on the unitarity triangle. 

For the first time the lattice spacing dependence of the matrix elements for such decays is studied. A similar analysis of $D$ mesons is described in Ref.~\cite{jims_talk}. 
Many of the simulation details are the same as those described in Ref.~\cite{fBpaper}, but a brief summary is given in Table~\ref{sim_details}.\footnote{Note that we have increased our statistics at $\beta = 6.1$ to 200 configurations.}

Our strategy is to study the $a$-dependence and perform the continuum extrapolations with a light quark (active and spectator) at strange. At $\beta = 5.7$ results are shown after the chiral extrapolation and an estimate of the remaining lattice spacing dependence is
made based on our study with strange light quarks. In a forthcoming paper we will also perform a chiral extrapolation at $\beta = 5.9$, allowing us to verify this $a$-dependence.
\begin{table} 
\vspace{-0.2in}
\begin{center}
\begin{tabular}{cccc}
\hline
\hline
$\beta$                 &6.1             &5.9             &5.7\\
\hline
Vol. 	                &$24^3\times 48$&$16^3\times 32$&$12^3\times 24$\\
\# cfgs.                &200 		&350            &300\\
$c_{\rm sw}$            &1.46	        &1.50           &1.57\\
$a^{-1}$ (GeV)          &2.62\err{8}{9}&1.80\err{5}{5} &1.16\err{3}{3}\\
$\kappa_h (=\kappa_b)$	&0.099	        &0.093	        &0.089\\
$\kappa_l$              &0.1373	        &0.1385         &0.1405\\
			&	        & 	        &0.1410\\
			& 	        & 	        &0.1415\\
			&               & 	        &0.1419\\
\hline
\hline
 			&               &       	&\\
\multicolumn{4}{l}{Table 1. Lattice details;
$a=a_{1P-1S}$.}
\end{tabular}
\end{center}
\label{sim_details} 
\vspace{-0.5in}
\end{table}
%
%
For both light and heavy quarks the action is the tadpole-improved, 
Sheikholeslami-Wohlert (SW) action with the plaquette value of $u_0$.
For heavy quarks it is interpreted in the Fermilab formalism~\cite{rev96}. 
The matrix elements and currents are calculated at tadpole-improved tree level as described in~\cite{rev96}.
The calculation of the current normalisation at one loop is  
underway~\cite{ask} and will be incorporated in our final results.
\section{The Calculation}
At each of the lattice spacings and for each momentum the required matrix element is 
extracted from the three-point correlation function
\begin{eqnarray*}
C^{3pt}(t,\vec{p},U)&\rightarrow& \frac{Z_\pi (\vec{p}_\pi )Z_B(\vec{p}_B)}{2E_\pi
(\vec{p}_\pi )2E_B(\vec{p}_B)}\times\\
&&\hspace{-0.5in}e^{-E_\pi t}e^{-E_B(T/2-t)}\langle \pi ,\vec{p}_\pi |V^\mu |B,\vec{p}_B
\rangle .
\end{eqnarray*}
The energies and amplitudes of the two-point functions were determined as described in Ref.~\cite{fBpaper}. In this simulation the $B$ meson is at rest and the light meson has momentum in lattice units of (0,0,0), (1,0,0), (1,1,0), (1,1,1) and (2,0,0).   
%
%

The spatial and temporal components of the matrix elements are interpolated in lattice 
momenta to fixed physical values, which, being matched at different lattice spacings, can 
be extrapolated to $a = 0$.
This introduces a dependence on the quantity used to set the scale, although it is found to be mild. Results are quoted using the 1P-1S splitting in charmonium to determine $a^{-1}$.
The linearly interpolated points are shown in Figure~1; the final result depends  only mildly on using a different interpolating function.
Typical linear continuum extrapolations of the matrix elements are in Figure~2. They show very little $a$-dependence however, the lattice spacing dependence does increase with $\vec{p}_\pi$, as expected.
\begin{figure} 
\vspace{-1.0cm}
\begin{center}
\leavevmode  
\epsfxsize=2.9in
\epsfysize=2.6in
\epsfbox{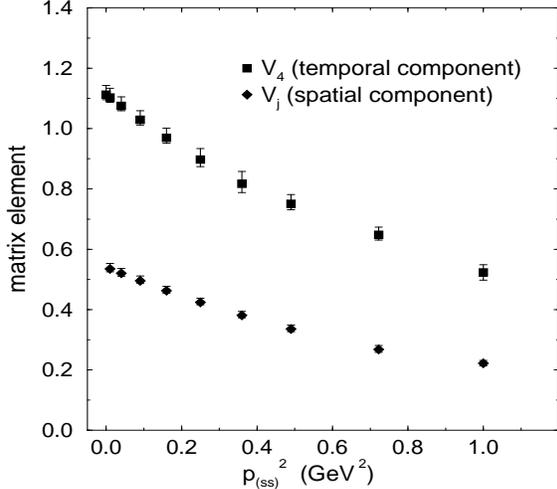}
\end{center}
\vspace{-0.58in}
\caption{Matrix elements at $\beta=5.7$ interpolated to $p\in\{$0.1, 0.2, 0.3, 0.4, 0.5, 0.6, 0.7, 0.85, 1.0$\}$ GeV.}
\end{figure}
\begin{figure} 
\vspace{-1.0cm}
\begin{center}
\leavevmode  
\epsfxsize=2.9in
\epsfysize=2.6in
\epsfbox{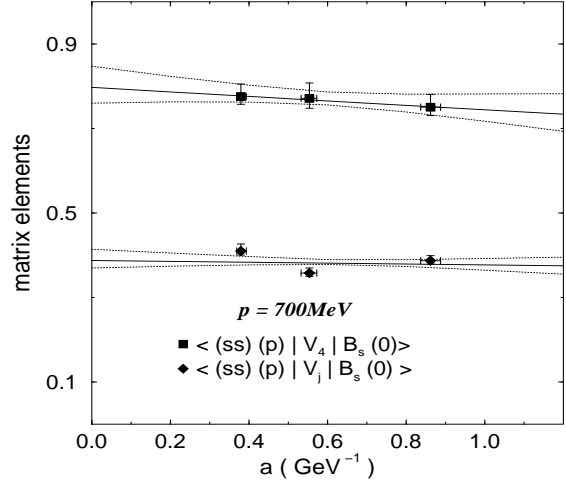}
\end{center}
\vspace{-0.58in}
\caption{Spatial and temporal components of the heavy-light matrix element at 
$\vec{p}_\pi = 700$ MeV extrapolated to the continuum limit.}
\end{figure}
%
%
%

At $\beta=5.7$ the matrix elements are extrapolated quadratically to the chiral limit. 
For momenta between $400$ and $850$ MeV the data fit a quadratic form very well and have good chi-squared. At $\vec{p}_\pi = 0$ 
the results become less reliable because the $B^\ast$ pole causes the extrapolation to rise rapidly with decreasing light quark mass. It would be useful to have still-lighter quarks to control this fit further, however exceptional configurations would have to be cured as in eg.~Ref.~\cite{MQA}.
At higher momenta the fits once more become less reliable due to large cutoff effects.
\section{Results}
Form factors are defined through
\begin{eqnarray*}
\langle\pi (p)|V^\mu |B(p^\prime)\rangle\hspace{-0.7em}&=&\hspace{-0.7em}f^+(q^2)\left [ p^\prime+p-\frac{m_B^2-m_\pi^2}{q^2}q\right ]^\mu \\
&+&\hspace{-0.7em}f^0(q^2)\frac{m_B^2-m_\pi^2}{q^2}q^\mu ,  
\end{eqnarray*}
with $q=p^\prime-p$.
One could then proceed to determine the form factors at $q^2 = 0$. 
However, we choose not to do so. 
Figure~3 illustrates firstly that our results for $f^0(q^2)$ and $f^+(q^2)$ 
agree with previous calculations using different approaches~\cite{UKQCD,LCSR} and secondly the huge range in $q^2$ 
over which lattice results must be extrapolated to reach $q^2=0$, thereby 
increasing the statistical error and introducing an unknown degree of model dependence 
in the final result. 
Instead~\cite{jns95} we focus on a quantity which can be determined without recourse to a $q^2$ extrapolation and which can be used with experimental data to determine $|V_{ub}|$. The 
differential decay rate is such a quantity, given by
\begin{equation}
\frac{d\Gamma}{d|\vec{p}_\pi|} = \frac{2m_B G_F^2|V_{ub}|^2}{24\pi^3}\frac{|\vec{p}_\pi|^
4}{E_\pi}\left | f^+(q^2)\right |^2 . 
\end{equation}
This is shown in Figure~4 where the 
dotted lines define the range in pion momentum for which we believe our lattice calculation is most reliable, and for which there are experimental results. Note that the $p_\pi = 0$ region is experimentally inaccessible since the event rate is zero here. 
\begin{figure} 
\vspace{-1.0cm}
\begin{center}
\leavevmode  
\epsfxsize=2.9in
\epsfysize=2.6in
\epsfbox{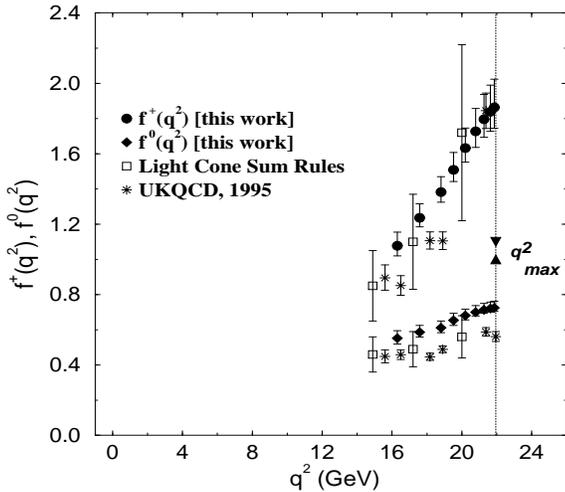}
\end{center}
\vspace{-0.58in}
\caption{The $q^2$-dependence of $f^0$ and $f^+$ (in the continuum limit) and a comparison with other calculations. Note that the lattice data only include statistical errors. }
\end{figure}
\begin{figure} 
\vspace{-1.0cm}
\begin{center}
\leavevmode  
\epsfxsize=2.9in
\epsfysize=2.6in
\epsfbox{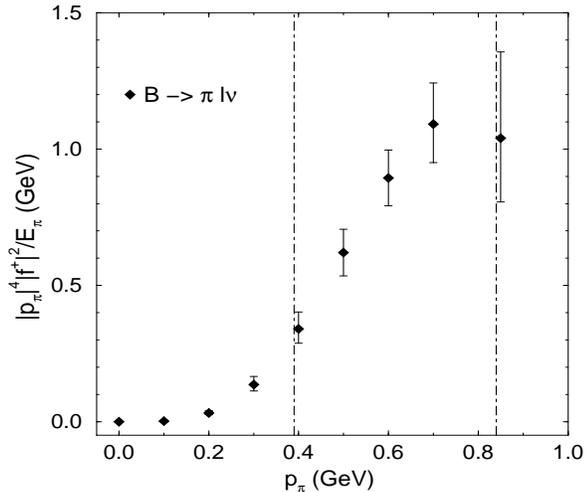}
\end{center}
\vspace{-0.58in}
\caption{The differential decay rate (less the momentum independent pre-factors) as a function of $\vec{p}_\pi$ at $\beta = 5.7$, chirally extrapolated.}
\end{figure}
%
%
%
\section{Systematic Errors}
Comparing data as in Figure~4 with 
the experimental measurements of exclusive branching ratios of $B$ mesons to pions,
eg.~in the range $400 \leq p_\pi \leq 850$ MeV, it is possible to determine $|V_{ub}|$. 
Estimates of the contributions to the theoretical error in $|V_{ub}|$ are tabulated below.
We expect to improve upon these in our final results.
\\
\\
\begin{tabular}{cc||cc}
Statistics &= 8\% & $a$-dependence &= 5\% \\
Pert$^n$. th.&= 5\% & Excited states &= 2\% \\
chiral extrap. &= 7\% &$m_Q$ tuning &= 1\% \\
\end{tabular}
\section{Conclusions}
Our preliminary results indicate that lattice spacing errors are under control and continuum extrapolations are possible. The form factors we extract agree with previous calculations for the range of $q^2$ where a lattice simulation is possible. We present an alternative way to use lattice data to determine $|V_{ub}|$, namely by calculating the partial widths, which should reduce statistical and systematic theoretical uncertainties~\cite{jns95}. 
%
\section*{Acknowledgements}
Fermilab is operated by Universities Research Association, Inc. for the U.S. Department of Energy.

\end{document}